\documentclass[twocolumn,showpacs,preprintnumbers,amssymb]{revtex4}
\usepackage{graphicx}
\usepackage{dcolumn}
\usepackage{amsmath}
\usepackage{amssymb}
\begin{document}
\preprint{APS/V2-15/10/02}

\title{Intermittency, scaling and the Fokker-Planck approach to\\
fluctuations of the solar wind bulk plasma parameters as seen by\\the WIND
spacecraft.}

\author{Bogdan Hnat}
 \email{hnat@astro.warwick.ac.uk}
\author{Sandra C. Chapman}
\author{George Rowlands}
 \affiliation{Physics Department, University of Warwick, Coventry, CV4 7AL, UK.}

\date{\today}

\begin{abstract}
The solar wind provides a natural laboratory for observations of MHD turbulence 
over extended temporal scales. Here, we apply a model independent method of
differencing and rescaling to identify self-similarity in the Probability
Density Functions (PDF) of fluctuations in solar wind bulk plasma parameters
as seen by the WIND spacecraft.
Whereas the fluctuations of speed $v$ and IMF magnitude $B$ are multi-fractal,
we find that the fluctuations in the ion density $\rho$, energy densities
$B^2$ and $\rho v^2$ as well as MHD-approximated Poynting flux $vB^2$ are
mono-scaling on the timescales up to $~26$ hours. The single curve, which we
find to describe the fluctuations PDF of all these quantities up to this
timescale, is non-Gaussian. We model this PDF with two approaches--
Fokker-Planck, for which we derive the transport coefficients and associated
Langevin equation, and the Castaing distribution that arises from a model for
the intermittent turbulent cascade.
\end{abstract}
\pacs{Valid PACS appear here}
\maketitle

\section{Introduction}
Statistical properties of velocity field fluctuations recorded in wind
tunnels and these obtained from solar wind observations exhibit striking
similarities \cite{carbone,veltri}. A unifying feature found in these
fluctuations is fractal or multi-fractal scaling.
The Probability Density Function (PDF), unlike power spectra that do not reveal
intermittency, show a clear departure from the Normal distribution when we
consider the difference in velocity on small spatial scales \cite{bohr,frisch}
while large scale features appear to be uncorrelated and converge toward a
Gaussian distribution.
These similarities suggest a common origin of the fluctuations in a
turbulent fluid and the solar wind. The approach is then to treat the solar
wind as an active highly nonlinear system with fluctuations arising in situ
in a manner similar to that of hydrodynamic turbulence \cite{cytu,goldstein,zelenyi,dobrowolny}.

Kolmogorov's K41 turbulence theory was based on the hypothesis that the
energy is transferred in the spectral domain at a constant rate through
local interaction within the inertial range. This energy cascade is
self-similar due to the lack of any characteristic spatial scale within the
inertial range itself.
These assumptions led Kolmogorov to his scaling law for the
moments of velocity structure functions \cite{frisch}:
$S_{\ell}^n=\left<|v(r+\ell)-v(r)|^n\right> \propto (\epsilon \ell)^{n/3}$,
where $n$ is the $n$-th moment, $\ell$ is a spatial scale and $\epsilon$
represents energy transfer rate.
Experimental results do not confirm this scaling, however, and modifications
to the theory include intermittency \cite{k62} by means of a randomly varying
energy transfer rate $\epsilon$. In this context, empirical models have been
widely used to approximate the shapes of fluctuation PDFs of data from wind
tunnels \cite{castaing} as well as the solar wind; see for example 
\cite{valvo,forman}.
The picture of turbulence emerging from these models is much more complex
then has been suggested by the original Kolmogorov theory. It requires a
multi-fractal phenomenology to be invoked as the self-similarity of the
cascade is broken by the introduction of the intermittency.

Recently, however, a new approach has emerged where the presence of
intermittency in the system coincides with statistical self-similarity,
rather than multi-fractality, in the fluctuations of selected quantities;
these also exhibit leptokurtic PDFs. An example of this {\it statistical
intermittency} was discussed in \cite{mantegna95}, where a L\'{e}vy
distribution was successfully fitted to the fluctuation PDFs of the price
index over the entire range of data. Such a distribution arises from the
statistically self-similar L\'{e}vy process also characterized by enhanced
(when compared with a Gaussian) probability of large events. Recently
\cite{hnat} reported similar self-similarity derived from the scaling of the
solar wind Interplanetary Magnetic Field (IMF) energy density fluctuations
calculated from the WIND spacecraft dataset.
\begin{figure}
\resizebox{\hsize}{!}{\includegraphics{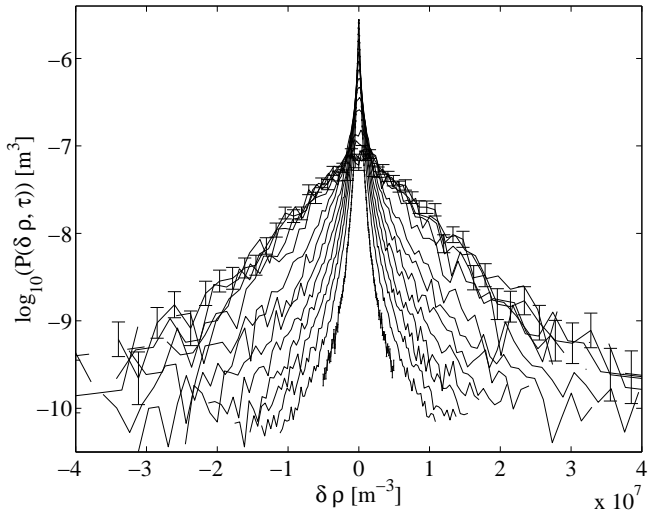}}
\caption{\label{fig1} Unscaled PDFs of the ion density fluctuations. Time lag
$\tau=2^k \times 46$s, where $k=0,1,2,..,14$. The standard deviation of the PDF
increases with $\tau$. The error bars on each bin within the PDF are
estimated assuming Gaussian statistics for the data within each bin.}
\end{figure}
Here, we apply a model-independent and generic PDF rescaling technique to extract the scaling properties of the solar wind fluctuations directly from
the data. The aim is to determine a set of plasma parameters that exhibit
statistical self-similarity and to verify the nature of the PDF for their
fluctuations.
We consider the following bulk plasma parameters: magnetic field magnitude
$B$, velocity magnitude $v$, ion density $\rho$, kinetic and magnetic energy
density ($\rho v^2$ and $B^2$) and Poynting flux approximated by $vB^2$.
Such an approximation of the Poynting flux assumes ideal MHD where
$\mathbf{E}=\mathbf{v}\times\mathbf{B}$. We find that the PDFs of fluctuations
in $\rho$, $B^2$, $\rho v^2$ and $vB^2$ exhibit mono-scaling for up to $10$
standard deviations, while $B$ and $v$ are clearly multi-fractal as found
previously \cite{burlaga,forman}.

The mono-scaling allows us to derive a Fokker-Planck equation that governs the
dynamics of the fluctuations' PDFs.
The Fokker-Planck approach provides a point of contact between the statistical
approach and the dynamical features of the system. This allows us to identify
the functional form of the space dependent diffusion coefficient that
describes the fluctuations of these quantities as well as to develop a
diffusion model for the shape of their PDFs.
We also consider a Castaing model where fluctuations are assumed to arise from
a varying energy transfer rate $\epsilon$ in the nonlinear energy cascade, with 
Gaussian distribution for $\ln(\epsilon)$.
\begin{figure}
\resizebox{\hsize}{!}{\includegraphics{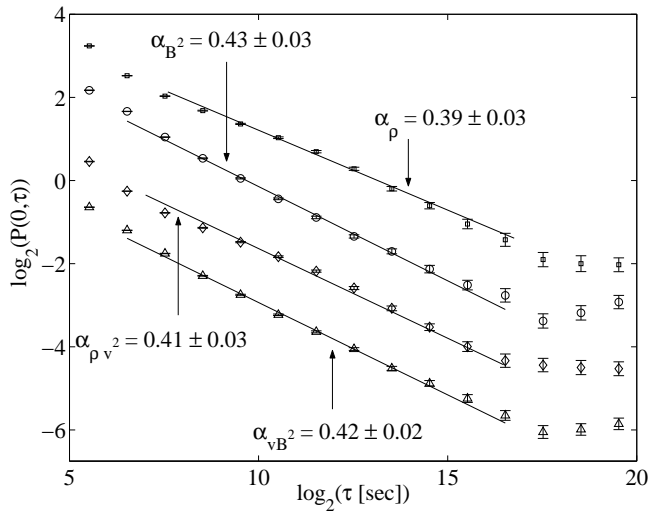}}
\caption{Scaling of the peaks $P(0,\tau)$ of the PDFs for all quantities under
investigation: $\circ$ corresponds to $\delta B^2$, $\square$ ion density
$\delta \rho$, $\diamond$ kinetic energy density $\delta (\rho v^2)$ and 
$\triangle$ Poynting flux component $\delta (vB^2)$. The plots have been offset
vertically for clarity. Errors are estimated as in Fig.~\ref{fig1}.}
\label{fig2}
\end{figure}
The paper is structured as follows: in section II we will describe the
dataset used for this study as well as the rescaling procedure. In section
III the results of the rescaling will be presented. Two possible models of
the fluctuations will be discussed in Section IV. Finally in Section V
we will summarize all results discussed throughout this paper.

\section{Data and Methods}
\subsection{The Dataset}
The solar wind is a supersonic, super-Alfv\'{e}nic flow of incompressible
and inhomogeneous plasma. The WIND spacecraft orbits the Earth-Sun L1 point
providing a set of in situ plasma parameters including magnetic field
measurements from the MFI experiment \cite{lepping} and the plasma parameters
from the SWE instrument \cite{ogilvie}.
The WIND solar wind magnetic field and key parameter database used here comprise
over $1.5$ million, $46$ second averaged samples from January 1995 to December
1998 inclusive. The selection criteria for solar wind data is given by the
component of the spacecraft position vector along the Earth-Sun line, $X>0$,
and the vector magnitude, $R>30$ RE.
The data set includes intervals of both slow and fast speed streams.
Similar to other satellite measurements, short gaps in the WIND data file
were present. To minimize the errors caused by such incomplete measurements we
omitted any intervals where the gap was larger than $2\%$. The original data
were not averaged nor detrended. The data are not sampled evenly but there are
two dominant sampling frequencies: $1/46$ Hz and $1/92$ Hz. We use sampling
frequency $f_s$ of $1/46$ as our base and treat other temporal resolutions as
gaps when the accuracy requires it ($\tau \leq 92$ seconds).

\subsection{Differencing and Rescaling Technique}
Let $x(t)$ represent the time series of the studied signal, in our case
magnetic field magnitude $B$, velocity magnitude $v$, ion density $\rho$,
kinetic energy density $\rho v^2$, magnetic field energy density $B^2$ or the
Poynting flux component approximated by $vB^2$.
A set of time series $\delta x(t,\tau)=x(t+\tau)-x(t)$ is obtained for each
value of the non-overlapping time lag $\tau$. The PDF $P(\delta x,\tau)$ is
then generated for each time series $\delta x(t,\tau)$. Fig.~\ref{fig1}
shows the set of such raw PDFs of the density fluctuations for time lags
between $46$ seconds and $\sim9$ days. 
\begin{figure}
\resizebox{\hsize}{!}{\includegraphics{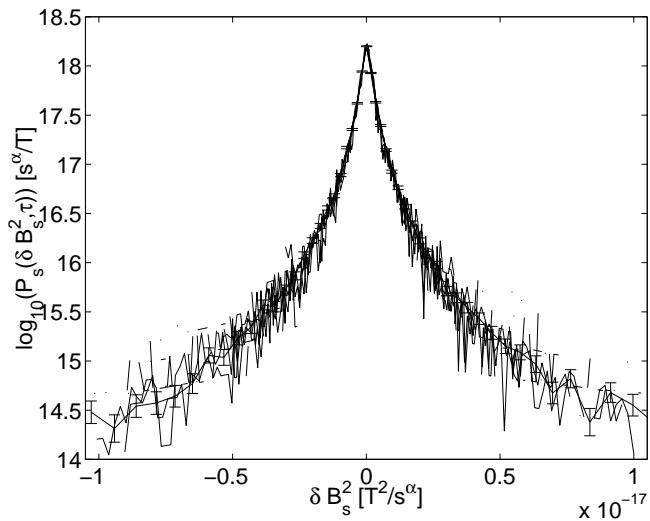}}
\caption{One parameter rescaling of the PDF for the fluctuations in the
magnetic field energy density $B^2$. The curves shown correspond to $\tau$
between $~2$ minutes and $~26$ hours. Error bars as in Fig.~\ref{fig1}.}
\label{fig3}
\end{figure}
A generic one parameter rescaling method \cite{hnat} is applied to these PDFs.
We extract the scaling index $\alpha$, with respect to $\tau$, directly from
the time series of the quantity $\delta x$.
Practically, obtaining the scaling exponent relies on the detection of
a power law, $P(0,\tau) \propto \tau^{-\alpha}$, for values of the raw PDF
peaks and time lag $\tau$. Fig.~\ref{fig2} shows the peaks $P(0,\tau)$ of the
unscaled PDFs plotted versus $\tau$ on log-log axes for the four bulk plasma
parameters. We see that the peaks of these PDFs are well described by a power
law $\tau^{-\alpha}$ for a range of $\tau$ up to $\sim26$ hours. We now take
$\alpha$ to be the scaling index and attempt to collapse all unscaled PDFs
$P(\delta x,\tau)$ onto a single curve $P_s(\delta x_s)$ using the following
change of variables:
\begin{equation}
P(\delta x,\tau)=\tau^{-\alpha} P_s(\delta x \tau^{-\alpha}).
\label{rescl}
\end{equation}
A self-similar Brownian walk with Gaussian PDFs on all temporal scales and
index $\alpha=1/2$ is a good example of the process where such collapse can be
observed (see e.g. \cite{sorn}). For experimental data, an approximate collapse
of PDFs is an indicator of a dominant self-similar trend in the time series,
i.e., this method may not be sensitive enough to detect multi-fractality that
could be present only during short time intervals.
One can treat the identification of the scaling exponent $\alpha$
and, as we will see, the non-Gaussian nature of the rescaled PDFs ($P_s$)
as a method for quantifying the intermittent character of the time series.
Another possible interpretation of the rescaling is to treat $P(\delta x,\tau)$
as the self-similar solution of the equation describing the PDF dynamics.
The mono-scaling of the fluctuations PDF, together with the finite value of
the samples' variance, indicates that a Fokker-Planck approach can be used to
express the dynamics of the unscaled PDF in time and with respect to the
coordinate $\delta x$ \cite{kampen}.
In section $IV$ we will use the Fokker-Planck equation to develop a dynamical
model for the fluctuations observed in the solar wind.
\begin{figure}
\resizebox{\hsize}{!}{\includegraphics{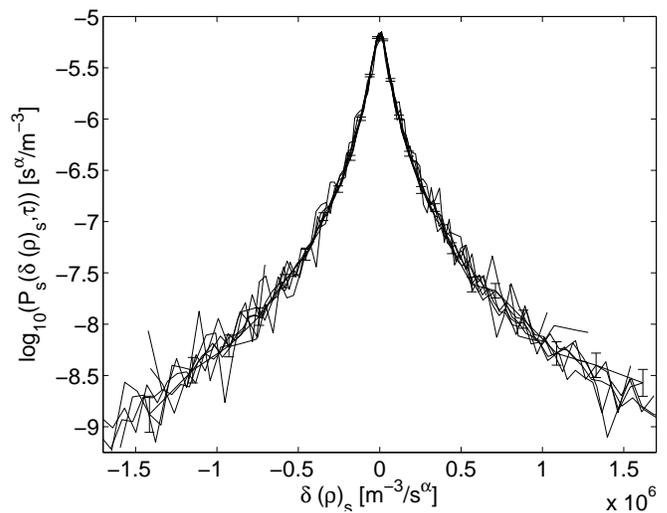}}
\caption{As in Fig.~\ref{fig3} for ion density fluctuations $\delta \rho$.} \label{fig4}
\end{figure}

Ideally, we use the peaks of the PDFs to obtain the scaling exponent $\alpha$,
as the peaks are statistically the most accurate parts of the distributions.
In certain cases, however, the peaks may not be the optimal statistical measure
for obtaining the scaling index. For example, the $B_z$ component of the solar
wind magnetic field is measured with an absolute accuracy of typically about
$0.1$ nT. Such discreteness in the time series introduces large errors in the
estimation of the peak values $P(0,\tau)$ and may not give a correct scaling.
However, if the PDFs rescale, we can in principle obtain the scaling exponent
from any point on the curve. We will illustrate this in the next section where
we obtain the rescaling index $\alpha$ from two points on the curve $P(0,\tau)$ and $P(\sigma,\tau)$.
\begin{figure}
\resizebox{\hsize}{!}{\includegraphics{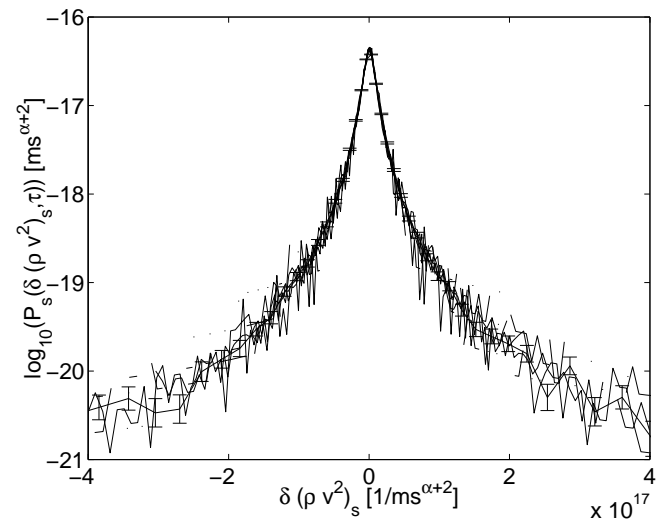}}
\caption{As in Fig.~\ref{fig3} for kinetic energy density fluctuations
$\delta (\rho v^2)$.}
\label{fig5}
\end{figure}

\section{PDF rescaling results}
\begin{figure}
\resizebox{\hsize}{!}{\includegraphics{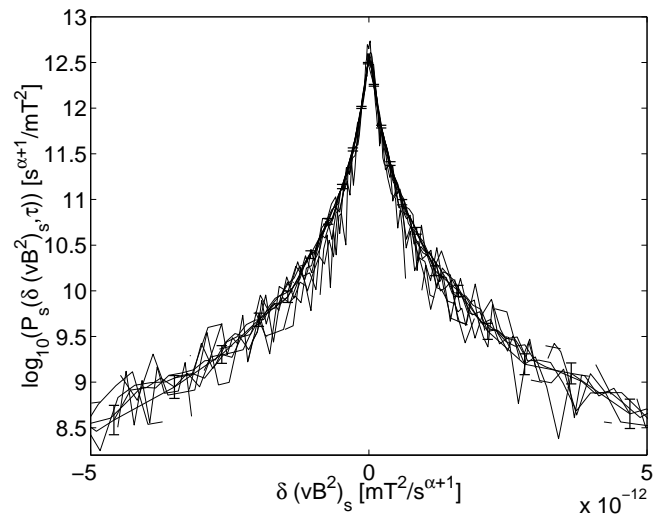}}
\caption{As in Fig.~\ref{fig3} for Poynting flux $\delta(vB^2)$.}
\label{fig6}
\end{figure}

We are now ready to present results of the rescaling procedure as applied
to the solar wind bulk plasma parameters. Fig.~\ref{fig1} shows the unscaled
(raw) PDF curves of the ion density data. These PDFs, like all others presented
in this section, were generated with the bin size decreasing linearly toward
the center of the distribution to improve the accuracy of the PDF for small
fluctuations. Although the entire range of data was used to create these PDFs
we truncated the plotted curves for $|\delta x| \geq 10 \sigma(\tau)$, where
$\sigma(\tau)$ is a standard deviation of the differenced time series for the
specific time lag $\tau$.
Fig.~\ref{fig2} then shows $P(0,\tau)$ plotted versus $\tau$ on log-log axes
for $\delta x=\delta(\rho)$, $\delta(\rho v^2)$, $\delta(B^2)$ and
$\delta(vB^2)$. Straight lines on such a plot suggest that the rescaling
(\ref{rescl}) holds at least for the peaks of the distributions. In Fig.
~\ref{fig2}, lines were fitted with $R^2$ goodness of fit for the range of
$\tau$ between $~2$ minutes and $~26$ hours, omitting points corresponding to the first two temporal scales as in these cases the sharp peaks of the PDFs
can not be well resolved. The lines suggest self-similarity persists up to
intervals of $\tau \approx 26$ hours. The slopes of these lines yield the
exponents $\alpha$ and these are summarized in Table \ref{tab1} along with the
values obtained from analogous plots of $P(\sigma(\tau),\tau)$ versus $\tau$
which show the same scale break and the same scaling exponent for
$\delta(\rho)$, $\delta(\rho v^2)$, $\delta(B^2)$ and $\delta(vB^2)$, to
within the estimated statistical error.
\begin{table}[b]
\begin{center}
\begin{tabular}{|p{1.25cm}||p{2.25cm}|p{2.25cm}|p{1cm}|p{1.1cm}|}
\hline
Quantity&$\alpha$ from $P(0,\tau)$&$\alpha$ from $P(\sigma,\tau)$&Approx.  $\tau_{max}$&PDF scales\\
\hline 
$\delta B$&$-0.47 \pm 0.02$&$-0.23 \pm 0.05$&$~26$ hrs&No\\
\hline
$\delta v$&$-0.52 \pm 0.05$&$-0.21 \pm 0.06$&$~26$ hrs&No\\
\hline
$\delta(B^2)$&$-0.43 \pm 0.03$&$-0.39 \pm 0.08$&$~26$ hrs&Yes\\
\hline
$\delta(\rho)$&$-0.39 \pm 0.03$&$-0.37 \pm 0.05$&$~26$ hrs&Yes\\
\hline
$\delta(\rho v^2)$&$-0.41 \pm 0.03$&$-0.35 \pm 0.05$&$~26$ hrs&Yes\\
\hline
$\delta(vB^2)$&$-0.42 \pm 0.02$&$-0.39 \pm 0.06$&$~26$ hrs&Yes\\
\hline
\end{tabular}
\caption{Scaling indices derived from $P(0,\tau)$ and P($\sigma,\tau)$ power
laws.}
\label{tab1}
\end{center}
\end{table}
Within this scaling range we now attempt to collapse each corresponding
unscaled PDF onto a single master curve using the scaling (\ref{rescl}).
Figs.~\ref{fig3}-\ref{fig6} show the result of the one parameter
rescaling applied to this unscaled PDF of fluctuations in $\rho$,
$\rho v^2$, $B^2$ and $vB^2$ respectively, for temporal scales up to
$\sim26$ hours. We see that the rescaling procedure (\ref{rescl}) using the
value of the exponent $\alpha$ of the peaks $P(0,\tau)$ shown in 
Fig.~\ref{fig2}, gives good collapse of each curve onto a single common
functional form for the entire range of the data. These rescaled PDFs are
leptokurtic rather than Gaussian and are thus strongly suggestive of an
underlying nonlinear process.

The fluctuations PDFs for all mono-scaling quantities investigated here are
nearly symmetric. This is in sharp contrast with the strong asymmetry of the
PDF of velocity fluctuations in hydrodynamic turbulence reported previously in
\cite{vainshtein,castaing}.
This asymmetry of the statistics for the velocity increments coincides with
the highly intermittent character of the flow and multi-fractal scaling of
these fluctuations.
We applied zeroth-order correlation functions, defined separately for the
positive and the negative branch of the PDF \cite{vainshtein}, to quantify
the asymmetry of fluctuations PDFs for the solar wind. This analysis was
performed using PDFs generated for $\tau \approx 12$ minutes (that is, within
the scaling region).
In the case of velocity increments we find that the negative moment is, on
average, $11\%$ lower compared to the positive one. On the other hand, the
quantities that we have found with self-similar increments do not have appreciable asymmetry. These show differences between negative and positive
moments of about $2-3\%$, which is however, well above the statistical error
of this procedure.

\begin{figure}
\resizebox{\hsize}{!}{\includegraphics{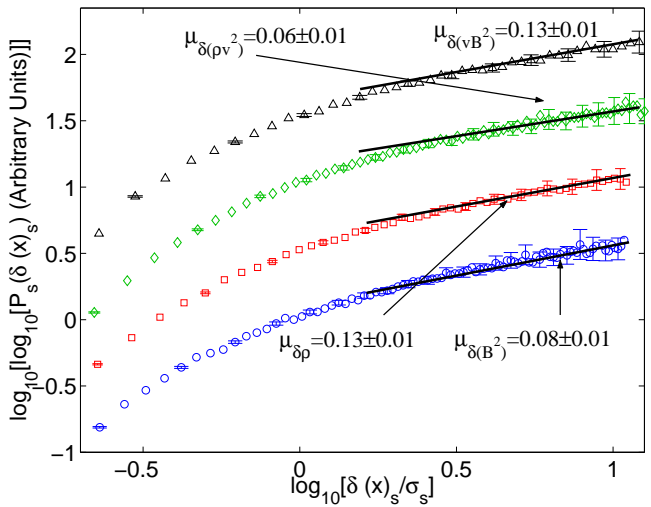}}
\caption{Positive tails of all self-similar PDFs for $\tau\approx 30$ minutes. Solid lines show linear fits obtained in the interval $[3\sigma,6\sigma]$ and extended to $[2\sigma,10\sigma]$ fluctuations interval.}
\label{fig7}
\end{figure}

It has been reported previously \cite{castaing} that the
PDFs obtained from hydrodynamic turbulence have exponential tails. These would
look linear on the linear-log plots that are used in this paper.
In the case of solar wind bulk plasma parameters we do not find such a clear
exponential cutoff region but rather see stretched exponential tails of the
form $P(|\delta x|)\sim\exp(-A|\delta x|^\mu)$. This is illustrated in
Fig.~\ref{fig7} where we plot $\log(\log(P(\delta x)))$ against
$\log(\delta x)$ for all positive fluctuations of the mono-scaling quantities. 
It can be seen that, as we move away from the peak, these curves converge
to lines and good fits can be obtained in the interval $[2\sigma,10\sigma]$, where $\sigma$ stands for the standard deviation.

\begin{figure}
\resizebox{\hsize}{!}{\includegraphics{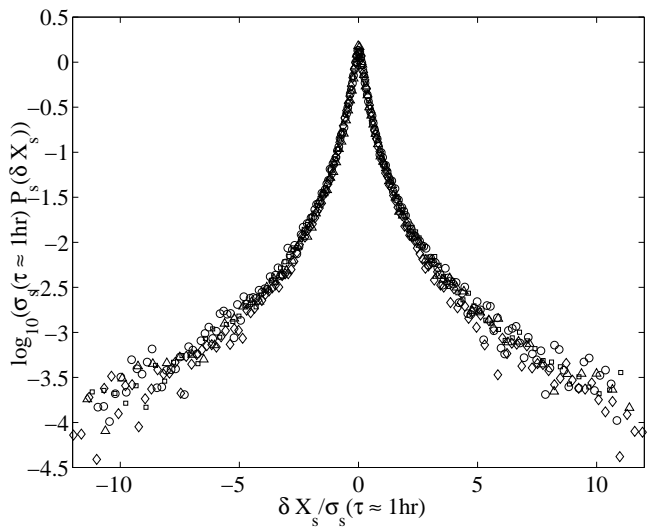}}
\caption{Direct comparison of the PDFs of fluctuations for all four 
quantities. $\circ$ corresponds to $\delta(B^2)$, $\square$ ion density
$\delta(\rho$), $\diamond$ kinetic energy density $\delta(\rho v^2)$ and
$\triangle$ Poynting flux component $\delta(vB^2)$.}
\label{fig8}
\end{figure}

We can now directly compare the functional form of these rescaled PDFs by
normalizing the curves and overlying them on the single plot for a particular
$\tau$ within the scaling range.
Fig.~\ref{fig8} shows these normalized PDFs $P_s(\delta x_s,\tau)$ for
$\delta x_s=\delta(\rho)_s$, $\delta(B^2)_s$, $\delta(\rho v^2)_s$,
$\delta(vB^2)_s$ and $\tau \approx 1$ hour overlaid on a single plot.
The $\delta x_s$ variable has been normalized to the rescaled standard
deviation $\sigma_s(\tau\approx 1hr)$ of $P_s$ and the values of the PDF has
been modified to keep probability constant in each case to facilitate this
comparison. These normalized PDFs have remarkably similar functional form
suggesting a shared process responsible for fluctuations in these four
plasma parameters on temporal scales up to $\tau_{max}\approx 26$ hours. 

\begin{figure}
\resizebox{\hsize}{!}{\includegraphics{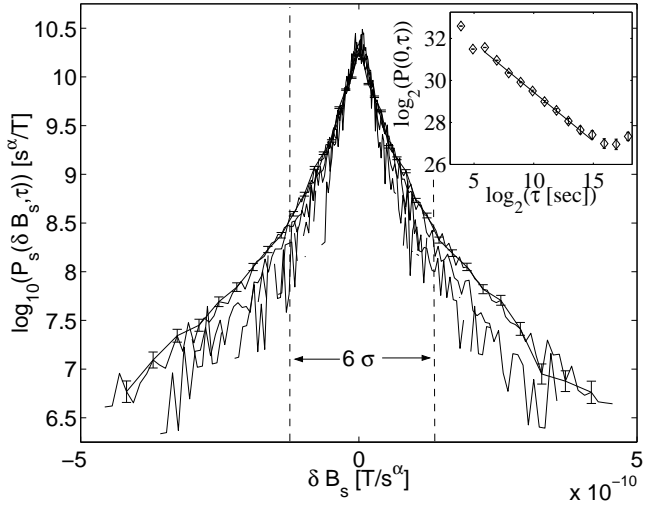}}
\caption{As in Fig.~\ref{fig3} for the solar wind magnetic filed magnitude
fluctuations.}
\label{fig9}
\end{figure}

\begin{figure}
\resizebox{\hsize}{!}{\includegraphics{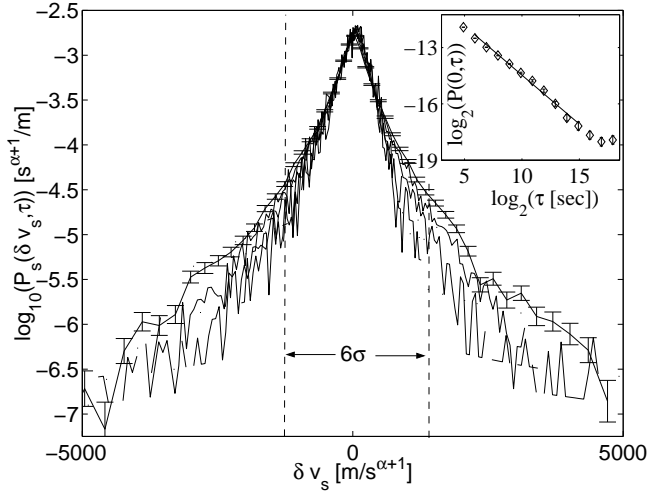}}
\caption{As in Fig.~\ref{fig3} for the solar wind velocity fluctuations.}
\label{fig10}
\end{figure}

It has been found previously \cite{burlaga} that the magnetic field magnitude
fluctuations are not self-similar but rather multi-fractal.
For such processes the scaling derived from $P(0,\tau)$ would not be expected
to rescale the entire PDF. To verify this we applied the rescaling procedure
for magnetic field magnitude differences $\delta B(t,\tau)=B(t+\tau)-B(t)$.
Fig.~\ref{fig9} shows the result of one parameter rescaling applied to
the PDFs of the magnetic field magnitude fluctuations. We see that the
scaling procedure is satisfactory only up to $\sim3$ standard deviations of
the original sample, despite the satisfactory scaling obtained for the peaks
$P(0,\tau)$ of the PDFs (see insert of the Fig.~\ref{fig9}). This confirms
the results of \cite{valvo} where a two parameter Castaing fit to values
within $3$ standard deviations of the original sample yields scaling in one
parameter and weak variation in the other.
Attempts to improve the collapse by using information in the tails (values
$|\delta B|>3\sigma$) would introduce a significant error in the estimation of
the scaling exponent $\alpha$. We found similar lack of scaling in the
fluctuations of the solar wind velocity magnitude and we show the rescaled PDF
in the Fig.~\ref{fig10}. We stress that the log-log plots of the PDF peaks
$P(0,\tau)$ show a linear region for both velocity and magnetic field magnitude
fluctuations (see insert in each figure). Their PDFs, however, do not collapse
onto a single curve when the rescaling (\ref{rescl}) is applied. This lack of
mono-scaling is evident when indices derived from $P(0,\tau)$ and these found
for $P(\sigma,\tau)$ are compared (see Table \ref{tab1}).

\section{Modelling the data}
The rescaling technique applied in the previous section indicates that,
for certain temporal scales, the PDFs of some bulk plasma parameters can be
collapsed onto a single master curve. The challenge now lays in developing physical models that can
describe the functional form of this curve.
Here we consider two approaches. The first is a statistical approach where
we assume that the fluctuations can be described by a stochastic Langevin
equation. The second method is to assume the fluctuations are the result
of the nonlinear energy cascade and derive the corresponding PDF form
for the rescaled PDFs (Castaing distribution) \cite{castaing}.

\subsection{Diffusion model}
\begin{figure}
\resizebox{\hsize}{!}{\includegraphics{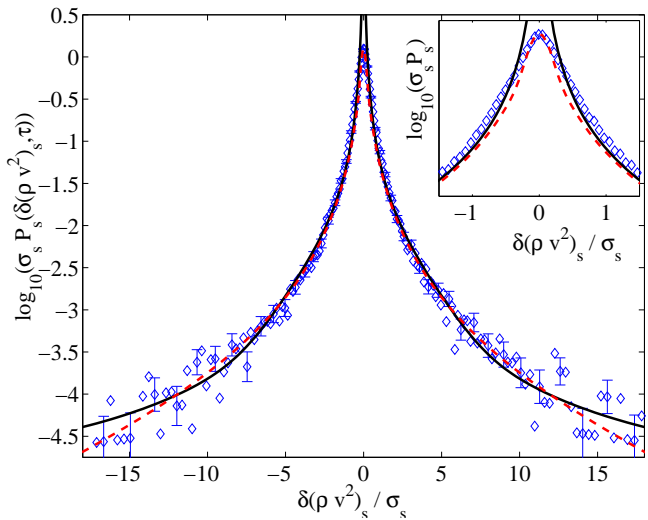}}
\caption{Example of the fit of the PDF functional form predicted by a
Fokker-Planck description (\ref{mdl2}) (solid line) and a Castaing model
(dash line) to the fluctuations PDF of the $\delta(\rho v^2)$ bulk parameter.}
\label{fig11}
\end{figure}

The Fokker-Planck (F-P) equation provides an important link between statistical
studies and the dynamical approach expressed by the Langevin equation
\cite{sorn}. In the most general form F-P can be written as:
\begin{equation}
\frac{\partial{P}}{\partial{\tau}}=
\nabla_{\delta x} (A(\delta x)P + B(\delta x)\nabla_{\delta x}P), 
\label{f-p}
\end{equation}
where $P \equiv P(\delta x,\tau)$ is a PDF for the differenced quantity
$\delta x$ that varies with time $\tau$, $A(\delta x)$ is the friction
coefficient and $B(\delta x)$ is related to a diffusion coefficient which we
allow to vary with $\delta x$.
For certain choices of $A(\delta x)$ and $B(\delta x)$, a class of self-similar
solutions of (\ref{f-p}) satisfies the rescaling relation given by 
(\ref{rescl}). This scaling is a direct consequence of the fact that the F-P
equation is invariant under the transformation
$\delta x \rightarrow \delta x \tau^{-\alpha}$ and $P \rightarrow P\tau^{\alpha}$.
It can be shown (see Appendix A) that equations (\ref{rescl}) and (\ref{f-p})
combined with power law scaling of the transport coefficients $A(\delta x)$
and $B(\delta x)$ lead to the following equation for the PDF:
\begin{equation}
\frac{\partial{P}}{\partial{\tau}}=\frac{\partial}{\partial{(\delta x)}}\left[(\delta x)^{1-1/\alpha}\left(a_0P + b_0 \delta x \frac{\partial{P}}{\partial{(\delta x)}}\right)\right],
\label{mdl1}
\end{equation}
where $a_0$ and $b_0$ are constants, $\alpha$ is the scaling index derived
from the data and $P(\delta x)$ and $\delta x$ are unscaled PDF and
fluctuations respectively. Written in this form equation (\ref{mdl1})
immediately allows us to identity the functional form of the diffusion
coefficient, namely $D(\delta x) \propto (\delta x)^{2-1/\alpha}$.
In Appendix A we show how (\ref{mdl1}) can also be expressed as:
\begin{equation}
\frac{b_0}{a_0}(\delta x_s)\frac{dP_s}{d(\delta x_s)}+P_s+\frac{\alpha}{a_0} (\delta x_s)^{\frac{1}{\alpha}}P_s = C.
\label{mdl1.5}
\end{equation}
The partial differential equation (\ref{mdl1.5}) can be solved analytically
and one arrives at the general solution in the form:
\begin{eqnarray}
P_s(\delta x_s)=\frac{a_0}{b_0}\frac{C}{|\delta x_s|^{a_0/b_0}} exp\left(-\frac{\alpha^2}{b_0}(\delta x_s)^{1/\alpha}\right)\nonumber \\ \times \int_0^{\delta x_s} \frac{exp\left(\frac{\alpha^2}{b_0}(\delta x_s')^{1/\alpha}\right)}{(\delta x_s')^{1-a_0/b_0}}d(\delta x_s') + k_0H(\delta x_s),
\label{mdl2}
\end{eqnarray}
where $k_0$ is a constant and $H(\delta x_s)$ is the homogeneous solution:
\begin{equation}
H(\delta x_s)=\frac{1}{(\delta x_s)^{a_0/b_0}} exp\left(-\frac{\alpha^2}{b_0}(\delta x_s)^{1/\alpha}\right).
\label{homsol}
\end{equation}
We then attempt to fit the predicted solution (\ref{mdl2}) to the normalized
rescaled PDFs. The results of such a fit for the fluctuations of the kinetic
energy density PDF is shown in Fig.~\ref{fig11} (solid line).
This fit is obtained with the following parameters $a_0/b_0=2.0$, $b_0=10$,
$C=0.00152$, $k_0=0.0625$ and $\alpha=0.41$ as derived from the rescaling
procedure.
We note that the figure is a semi-log plot and thus emphasizes the tails of the 
distribution - for a different value of the ratio $a_0/b_0$ the fit around
the smallest fluctuations could be improved. Equation (\ref{mdl2}) can not,
however, properly model the smallest fluctuations as it diverges for
$\delta x_s \rightarrow 0$.

Let us now assume that a Langevin equation in the form
\begin{equation}
\frac{d(\delta x)}{dt}=\beta(\delta x)+\gamma(\delta x)\xi(t)
\label{langevin}
\end{equation}
can describe the dynamics of the fluctuations. In (\ref{langevin}) the
random variable $\xi(t)$ is assumed to be $\delta$-correlated, i.e.,
\begin{equation}
<\xi(t)\xi(t+\tau)>=\sigma^2 \delta(\tau).
\label{dcorr}
\end{equation}
This condition is fulfilled in the data analysis by forming each time series
$\delta x(t,\tau)$ with non-overlapping time intervals $\tau$ and was also
verified by computing the autocorrelation function of the differenced time
series. Introducing a new variable
$z=\int_0^{\delta x} 1/\gamma(\delta x') d(\delta x')$, equation (\ref{langevin})
can be written as:
\begin{equation}
\frac{dz}{dt} = \frac{\beta(z)}{\gamma(z)} + \xi(t).
\label{langevin1}
\end{equation}
One can immediately obtain a F-P equation that corresponds to the Langevin
equation (\ref{langevin1}) \cite{kampen}. We can then compare this F-P
equation with that given by (\ref{mdl1}) to express coefficients
$\beta(\delta x)$ and $\gamma(\delta x)$ in terms of $a_0$ and $b_0$
(see Appendix B). Defining $D_0=<\xi^2(t)>/2$ we obtain:
\begin{equation}
\gamma(\delta x)=\sqrt{\frac{b_0}{D_0}}(\delta x)^{1-\frac{1}{2\alpha}},
\label{lngb}
\end{equation}
and
\begin{equation}
\beta(\delta x)=[b_0(1-\frac{1}{2\alpha})-a_0](\delta x)^{1-\frac{1}{\alpha}}.
\label{lngg}
\end{equation}
Equation (\ref{langevin}) together with definitions of its coefficients
(\ref{lngb}) and (\ref{lngg}) constitutes a dynamical model for the
fluctuations in the solar wind quantities. From (\ref{lngb}) and
(\ref{lngg}), we see that the diffusion of the PDF of fluctuations in the
solar wind is of comparable strength to the advection ($a_0/b_0 \approx 2$).
We stress that the advection and diffusion processes that we discuss here
are of the probability in parameter space for fluctuations and do not refer
to the integrated quantities.

\subsection{Castaing model}
We now, for comparison, consider a model motivated directly by a cascade in
energy, due to Castaing. This empirical model was developed
for the spatial velocity fluctuations recorded from controlled experiments
in wind tunnels \cite{castaing,vanatta} and has been applied to the solar wind
data \cite{valvo,forman}. The underlying idea of this approach is that, for
constant energy transfer rate between spatial scales, all quantities should
exhibit a Gaussian distribution of fluctuations.
The intermittency is then introduced to the PDF through the fluctuations of
the variance $\sigma$ of that Gaussian distribution. A log-normal
distribution is assumed for the variance $\sigma$:
\begin{equation}
Q(\sigma)=\frac{1}{\sqrt{2\pi}\lambda}exp\left(-\frac{ln^2(\sigma/\sigma_0)}{2\lambda^2}\right) d(ln(\sigma)),
\label{lognrml}
\end{equation}
where $\sigma_0$ is the most probable variance of the fluctuations and
$\lambda$ is the variance of $ln(\sigma)$. Combining these two hypothesis
Castaing proposed the following functional form for the observed PDF:
\begin{equation}
P_{\lambda}(\delta x)= \frac{1}{2\pi\lambda} \int_0^{\infty}exp\left(-\frac{(\delta x)^2}{2\sigma^2}\right) exp\left(-\frac{ln^2(\sigma/\sigma_0)}{2\lambda^2}\right) \frac{d\sigma}{\sigma^2}.
\label{castaing}
\end{equation}
The dashed line in the Fig.~\ref{fig11} shows the Castaing curve
fitted with parameters $\lambda=1.275$ and $\sigma_0=0.225$ to the
$\delta(\rho v^2)$ PDF.

We can now compare the rescaled PDFs with both F-P and Castaing predicted
curves which are shown in Fig.~\ref{fig11}. We can see from the figure that
both models provide an adequate fit to the $\delta(\rho v^2)_s$ PDF, and hence
will also describe the PDF of other scaling bulk plasma parameters.
Both curves, however, fall significantly below observed PDF values for
$|\delta(\rho v^2)_s| \leq 2$, although the Castaing distribution fits the peak
of the PDF reasonably well (see insert in Fig.~\ref{fig11}). This departure
from the experimental PDF, in the case of the Castaing distribution, may
reflect the difference between hydrodynamics and MHD turbulence.

\section{Summary}
In this paper we have applied a generic PDF rescaling method to fluctuations 
in the solar wind bulk plasma parameters. We find that, consistent with
previous work, magnetic field and velocity magnitude fluctuations are
multi-fractal whereas the PDFs of fluctuations in $B^2$, $\rho$, $\rho v^2$
and $vB^2$ can be rescaled with just one parameter for temporal scales up to
$\sim26$ hours.
The presence of intermittency in the plasma flow is manifested in these
quantities simply by the leptokurtic nature of their fluctuation PDFs, which
show increased probability of large fluctuations compared to that of the Normal
distribution. Fluctuations on large temporal scales, $\tau>26$ hours are uncorrelated in that their PDFs converge toward a Gaussian distribution.
The fact that all quantities share the same PDF, to within errors, is also
strongly suggestive of a single underlying process. This is also supported by
the similar values of the scaling exponents.

The simple scaling properties that we have found allow us to develop a
Fokker-Planck approach which provides a functional form of the rescaled PDFs as
well as a Langevin equation for the dynamics of the observed fluctuations.
The model shows that both advective and diffusive terms need to be invoked 
to describe the dynamics of the fluctuations. The calculated diffusion
coefficient is of the form $D(x_s) \propto (\delta x_s)^{2-1/\alpha}$.
We obtained a good fit of the model to our rescaled PDFs over at least $10$
standard deviations. We also examined a Castaing model for turbulence and found
a set of fit parameters for which both the Castaing distribution and our
diffusion model have nearly identical form. Since both the F-P model and the
Castaing distribution fit our rescaled PDFs we conclude that their moments
should exhibit same variation with time lag $\tau$.

\section{Acknowledgment} 
S. C. Chapman and B. Hnat acknowledge support from the PPARC and G. Rowlands
from the Leverhulme Trust. We thank N. W. Watkins and M. P. Freeman for advice 
concerning the post processing of the WIND data. We also thank R.P Lepping and
K. Ogilvie for provision of data from the NASA WIND spacecraft.

\appendix
\section{}
Let $P(\delta x,\tau)$ be a homogeneous function that satisfies scaling
(\ref{rescl}). Our aim is to find functional form of the coefficients
$A(\delta x)$ and $B(\delta x)$ for which $P(\delta x,\tau)$ is a solution
of a F-P equation (\ref{f-p}). Using (\ref{rescl}) we can now rewrite
(\ref{f-p}) to read:
\begin{eqnarray}
&&-\frac{\alpha}{t^{\alpha+1}}\left(P_s+\delta x_s \frac{dP_s}{d(\delta x_s)}\right)=\frac{P_s}{t^{\alpha}}\frac{dA(\delta x)}{d(\delta x)} + \frac{A(\delta x)}{t^{2\alpha}}\frac{dP_s}{d(\delta x_s)} \nonumber \\
&&+ \frac{1}{t^{2\alpha}}\frac{dB(\delta x)}{d(\delta x)}\frac{dP_s}{d(\delta x_s)} + \frac{B(\delta x)}{t^{3\alpha}}\frac{dP_s}{d(\delta x_s)}.
\label{app1}
\end{eqnarray}
If all terms in the rhs of (\ref{app1}) are to contribute and for $P(\delta x_s)$
to remain a function of $\delta x_s$ only we must have:
\begin{equation}
\frac{A(\delta x)}{t^{\alpha-1}}=a(\delta x_s) \quad \textrm{and} \quad \frac{B(\delta x)}{t^{2\alpha-1}}=b(\delta x_s).
\label{app2}
\end{equation}
Both $A(\delta x)$ and $B(\delta x)$ must then be of form:
\begin{equation}
A(\delta x)=a_0(\delta x)^{\eta} \quad \textrm{and} \quad B(\delta x)=b_0(\delta x)^{\nu},
\label{app3}
\end{equation}
where $a_0$ and $b_0$ are constants. Changing variables to the rescaled
$\delta x_s$ and substituting (\ref{app3}) into (\ref{app2}) we express
exponents $\eta$ and $\nu$ in terms of the rescaling index $\alpha$ derived
from the data. We then obtain:
\begin{equation}
\eta=1-\frac{1}{\alpha} \quad \textrm{and} \quad \nu=2-\frac{1}{\alpha},
\label{app4}
\end{equation}
which allows to write the final power law form of $A(\delta x)$ and
$B(\delta x)$:
\begin{equation}
A(\delta x)=a_0(\delta x)^{1-\frac{1}{\alpha}} \quad \textrm{and} \quad B(\delta x)=b_0(\delta x)^{2-\frac{1}{\alpha}}.
\label{app5}
\end{equation}
Substituting these expressions into F-P equation (\ref{f-p}) we obtain
(\ref{mdl1}) from Section $4$. Using these results the term
$\frac{dA(\delta x)}{d(\delta x)}$ on the rhs of (\ref{app1}), for example,
becomes:
\begin{equation}
\frac{dA(\delta x)}{d(\delta x)}=\left(1-\frac{1}{\alpha}\right)a_0(\delta x)^{-\frac{1}{\alpha}}.
\label{app6}
\end{equation}
Performing similar algebra on all terms in (\ref{app1}) we arrive to
equation:
\begin{equation}
-\alpha\frac{d(\delta x_s P_s)}{d(\delta x_s)}=\frac{d}{d(\delta x_s)} \left[(\delta x_s)^{1-\frac{1}{\alpha}}\left(a_0P_s+b_0(\delta x_s)\frac{dP_s}{d(\delta x_s)}\right)\right].
\label{app7}
\end{equation}
Integrated once we obtain equation (\ref{mdl1.5})
\begin{equation}
\frac{b_0}{a_0}(\delta x_s)\frac{dP_s}{d(\delta x_s)}+ P_s+\frac{\alpha}{a_0} (\delta x_s)^{\frac{1}{\alpha}}dP_s = C,
\label{app8}
\end{equation}
where $C$ is the constant of integration.

\section{}
Consider the following Langevin type of equation:
\begin{equation}
\frac{d(\delta x)}{dt}=\beta(\delta x) + \gamma(\delta x)\xi(t),
\label{app9}
\end{equation}
where the random variable $\xi(t)$ is assumed to be $\delta$-correlated, i.e.,
\begin{equation}
<\xi(t)\xi(t+\tau)>=\sigma^2 \delta(\tau).
\label{app10}
\end{equation}
Introducing a new variable $z=\int_0^{\delta x} 1/\gamma(\delta x')d(\delta x')$,
equation (\ref{app9}) can be written as:
\begin{equation}
\frac{dz}{dt} = \Gamma(z)+\xi(t),\quad \textrm{where} \quad \Gamma(z)=\frac{\beta(z)}{\gamma(z)}.
\label{app11}
\end{equation}
One can immediately obtain a F-P equation that corresponds to
the Langevin equation (\ref{app11}) and reads:
\begin{equation}
\frac{\partial{P(z,\tau)}}{\partial{\tau}}+\frac{\partial}{\partial{z}}\left(\Gamma(z)P(z,\tau)\right)=D_0 \frac{\partial^2{P(z,\tau)}}{\partial^2{z}},
\label{app12}
\end{equation}
where $D_0=\sigma^2/2$. The probability is an invariant of the variable
change so that $P(\delta x)d(\delta x)=P(z)dz$ and we can then rewrite
(\ref{app12}) for $P(\delta x,\tau)$:
\begin{equation}
\frac{\partial{P}}{\partial{\tau}}=\frac{\partial}{\partial{(\delta x)}}\left[\left(D_0\gamma(\delta x)\frac{d\gamma(\delta x)}{d(\delta x)}- \beta(\delta x)\right)P+D_0\gamma^2 \frac{\partial{P}}{\partial{(\delta x)}}\right].
\label{app13}
\end{equation}
Comparing (\ref{app13}) with the F-P equation (\ref{mdl1}) we can identify:
\begin{equation}
D_0 \gamma^2 = (\delta x)^{1-\frac{1}{\alpha}} b_0 \delta x,
\label{app14}
\end{equation}
and then we must demand that:
\begin{equation}
\frac{D_0}{2}\frac{d\gamma^2(\delta x)}{d(\delta x)^2}-\beta(\delta x)=a_0 (\delta x)^{1-\frac{1}{\alpha}}.
\label{app15}
\end{equation}
In summary we have shown that the F-P equation given by (\ref{mdl1}) is
equivalent to the stochastic Langevin equation (\ref{langevin}) where
coefficients $\beta$ and $\gamma$ are given by:
\begin{equation}
\gamma=\sqrt{\frac{b_0}{D_0}}(\delta x)^{1-\frac{1}{2\alpha}},
\label{app16}
\end{equation}
and
\begin{equation}
\beta=\left[b_0\left(1-\frac{1}{2\alpha}\right)-a_0\right](\delta x)^{1-\frac{1}{\alpha}}.
\label{app17}
\end{equation}

\end{document}